\documentclass{revtex4}
\usepackage{graphicx,amsmath,amssymb,mathrsfs}
\usepackage{epsfig}

\begin{document}

\title{Cosmological flux noise and measured noise power spectra in SQUIDs}


\author{Christian Beck}

\affiliation{
School of Mathematical Sciences, Queen Mary University of London, Mile End Road, London E1 4NS, UK}

\begin{abstract}

{\bf The understanding of the origin of $1/f$ magnetic flux noise commonly observed in superconducting devices such as SQUIDs and
qubits is still a major unsolved puzzle.
Here we discuss the possibility
that a significant part
of the observed low-frequency flux noise measured in these devices 
is ultimately seeded by cosmological fluctuations.
We consider a theory where a primordial flux noise field left over
in unchanged form from an early inflationary or quantum gravity epoch of the universe
intrinsically influences the phase difference in SQUIDs and qubits.
The perturbation seeds generated by this field can explain in a quantitatively
correct way the form and amplitude of measured low-frequency flux noise spectra in SQUID devices if one takes as a source of fluctuations the
primordial power spectrum of curvature fluctuations as measured by the Planck collaboration.
Our theoretical predictions are in excellent agreement with recent low-frequency flux noise measurements of various experimental groups.
Magnetic flux noise, so far mainly considered as a nuisance for
 electronic devices, may thus contain valuable information about fluctuation spectra in the very early universe.}

\end{abstract}

\maketitle


\section*{Introduction}

The origin of $1/f$ noise in superconducting devices such as SQUIDs and qubits has remained
an unsolved puzzle over the past 30 years \cite{koch-first,wellstood,koch,martinis1,martinis2,sendel,bylander,clarke,anton,giazotto,wang}.
This noise limits the coherence time of superconducting qubits.
In contrast to other types of noises, it is notoriously difficult to construct
a plausible model of $1/f$ noise that is well-motivated on physical
grounds. For superconducting devices such as qubits and SQUIDs,
very precise measurements of the flux noise spectrum have recently become available,
both in the low-frequency region ($10^{-4}...10^{-1}$ Hz)\cite{martinis1, martinis2} as well as in the high
frequency region ($0.2...20$ MHz)\cite{bylander}. Still a fully convincing theory of the origin of the noise, in particular
in the low-frequency region, has not been achieved so far. Note that `noise' with a frequency of order $10^{-4}$ Hz
as measured in \cite{martinis2}
corresponds to a stochastic process that changes on a time scale of hours, which is difficult
to realize on an atomic or molecular level.

A useful effective model discussed in
\cite{koch,martinis2,clarke} is to attribute the $1/f$ flux noise to fluctuating spins of localized surface electrons,
assuming a very broad spectrum of local relaxation times. However,
the areal density of spins necessary to fit the observed typical magnitude of
the flux noise ($5 \cdot 10^{17}  m^{-2}$) is much higher than one would
normally expect for the materials considered \cite{martinis2}. Moreover, recent measurements of Anton
et al.\ \cite{clarke} cannot be explained with the assumption of independent surface
spins, one needs to assume clustered collective behavior of many spins.
While some experimental and theoretical progress has been made in the
past years on the (very weak) dependence of how the flux noise couples into
the measuring device as a function of its shape and other parameters \cite{koch,martinis1,martinis2,giazotto,clarke}, the deeper reason for the {\em a priori} origin of the magnetic flux noise is still
not 
understood, in particular in the low-frequency region $f<1$Hz, where it is
most intensive.

This has lead to a search for alternative explanations of the flux noise,
pointing towards other candidate sources in different areas of science. For example, a recent attempt
of Wang et al. \cite{wang}
relates some of the flux noise to absorbed oxygen molecules on the surface of the SQUID.
If this is true, then removing oxygen adsorbates from the surface of SQUIDs
would substantially reduce the flux noise amplitude, a fact that could be experimentally
tested in the future. It is likely that the ultimate theory of magnetic flux noise in SQUIDs
will point to a combination of many effects, some of them more fundamental than others.

In this paper we propose a new seed mechanism for the generation of flux noise
in SQUIDs at a fundamental level. Our theory is in excellent agreement with
experimental observations and
goes to a much deeper level of what the ultimate source of the
flux noise is, and why it is hard to shield
and avoid this noise at all. We propose that a significant part of the flux noise at low
frequencies is produced by cosmological seeds.
We will relate the intrinsic source to
the power spectrum of primordial density fluctuations in the early universe \cite{planck-old,planck,wmap},
conserved to the current time by a suitable cosmological field whose properties will be described in detail.
The primordial fluctuations are usually assumed to have been generated by quantum fluctuations
of the inflaton field during cosmological inflation. They can be
conserved to the current time in terms of misalignment angle fluctuations
of a very light frozen-in field that is a relict of the inflationary or quantum gravity phase
of the universe.

We will show that misalignment fluctuations can create
flux fluctuations.
When the Earth moves through the cosmologically generated pattern of small perturbations
of the misalignment angle, mirror fluctuations are induced for the phase difference
of the measuring Josephson junction.
The experimental consequence is $1/f$ flux noise, which, as we will show,
has the correct order of magnitude to explain the observed experimental data in the low-frequency region \cite{martinis1,martinis2,sendel}.
Surface effects, e.g.\ localized electrons or oxygen molecules \cite{wang}, are not in contradiction
to this theory, rather, they further modify
the cosmological seed signal at higher frequencies ($f > 1$ Hz).
In the low frequency region ($f\leq 1$ Hz),
we obtain
excellent quantitative agreement with experimentally measured flux noise spectra
without fitting any parameter.

Our proposed explanation of the flux noise falls into the general category
of experiments that test for tiny measurable fluctuations generated by the Earth moving relative to a given cosmic background field (see e.g.\
\cite{rapisarda} for another recent suggestion based on laser interferometry
and a movement of the Earth relative to the cosmic microwave background). If successful, these types of measurements could
open up a new experimental `window' to obtain information on the state of the universe at earliest times.


\section*{Results}

\subsection*{Theoretical prediction of a cosmological flux noise power spectrum}

The theory developed in this paper gives a
%
concrete prediction for the primary flux noise
power spectrum generated
in a SQUID due to cosmological fluctuations:

\begin{equation}
S_\Phi (f) = \frac{\theta_1^2}{16 \pi^2} \Phi_0^2 P(k)|_{k=f/v} \frac{1}{f}
\end{equation}
Here $P(k)$ is the primordial power spectrum of cosmological density fluctuations,
as generated e.g. in inflationary models, and $v$ is the velocity of the Earth
relative to the cosmic field background. $\Phi_0=h/2e$ denotes
the flux quantum. The angle $\theta_1\in [-\pi, \pi]$
denotes the initial value of the misalignment angle of the frozen-in cosmological field
that conserves the power spectrum to the current time.
Taking for $P(k)$ the primordial power spectrum of scalar perturbations
that is measured by the Planck satellite
\cite{planck-old,planck},
\begin{equation}
P(k)= \Delta_R^2 (k)= A_s \cdot \left( \frac{k}{k^*} \right)^{n_s -1}, \label{444}
\end{equation}
where $A_s=(2.14 \pm 0.05) \cdot 10^{-9}, n_s=0.968 \pm 0.006$, and $k^*=0.05$ Mpc$\,^{-1}$,
we get the concrete prediction
\begin{equation}
S_\Phi (f) = \frac{\theta_1^2\Phi_0^2 A_s}{16 \pi^2}\left( \frac{f}{vk^*} \right)^{n_s-1} \frac{1}{f}. \label{spectrum}
\end{equation}
This generates flux noise with an $1/f^{2-n_s}=1/f^\alpha \approx 1/f^{1.04}$ power spectrum. For the squared amplitude of this
noise we obtain at $f=1$ Hz, assuming $v \approx 368$ km/s (the velocity of the Earth relative to the reference frame
set by the cosmic microwave background)
\begin{equation}
 \left( \frac{\delta \Phi}{\Phi_0} \right)^2  = \frac{A_s}{16 \pi^2} \left( \frac{\mbox{1Hz}}{vk^*}\right)^{n_s-1}\theta_1^2 = (3.54 \pm 0.79) \cdot 10^{-12}\cdot \theta_1^2. \label{13}
\end{equation}
The error bars for the above numerical prediction are dominated by the precision by which the exponent $n_s$ is known
(we used the value
$n_s=0.968 \pm 0.006$ provided by the Planck collaboration in \cite{planck}). The dependence on the velocity $v$ in the above formula is very weak
because $n_s$ is close to 1. Hence uncertainties in the knowledge of $v$ induce only minor numerical differences.
For example, changing the velocity $v \approx 0.001c \to c$ by a factor 1000, the amplitude of the predicted flux noise
increases just by 11$\%$.

There is no {\em a priori} way to predict the initial value $\theta_1$
of the cosmological field angle, which arises due to spontaneous symmetry breaking
at the Planck scale. Still its order of magnitude
can be estimated by assuming that every value of
$\theta_1\in [-\pi, \pi]$ is equally likely.
By taking the uniform average, one obtains the average squared value
\begin{equation}
\bar{\theta_1^2} = \frac{1}{2\pi} \int_{-\pi}^\pi \theta_1^2 d\theta_1= \frac{\pi^2}{3}. \label{uni-ave}
\end{equation}
Putting this into eq.~(\ref{13}) one obtains the concrete numerical prediction
\begin{equation}
\frac{\delta \Phi}{\Phi_0}|_{ave}=(3.41 \pm 0.40) \cdot10^{-6} \label{888}
\end{equation}
at $f=$ 1 Hz. Once again let us mention that this is our prediction of the {\em primary}
flux noise power spectrum in SQUIDs as generated by cosmological effects.
This is then further modified by non-universal effects in a given physical
realization of a SQUID, which
depend (weakly) on dimensions of the SQUID and material parameters, in particular
in the high-frequency region $f>1$ Hz. On the other hand,
in the low frequency region $f \leq 1$ Hz, if a suitable experiment sensitive to
these low frequencies is performed, then
the measured magnetic flux noise spectrum in the SQUID is expected
to be close to the 
primary form as generated by cosmological seeds.


\subsection*{Comparison with experimental data}

Our predicted noise strength (\ref{888})
as well as the entire form of the spectrum is in excellent agreement with experimental results.
Let us first discuss the seminal flux noise measurements of Bialczak et al. \cite{martinis2} that
for the first time reached the low-frequency
region $10^{-5}$ Hz $<f<10^{-1}$ Hz.
Fig.~1 shows these data together with our theoretical prediction given by eq.~(\ref{spectrum}),
using for $\theta_1^2$ the cosmological average value $\bar{\theta_1^2}=\pi^2/3$.
\begin{figure}
\includegraphics[width=8cm]{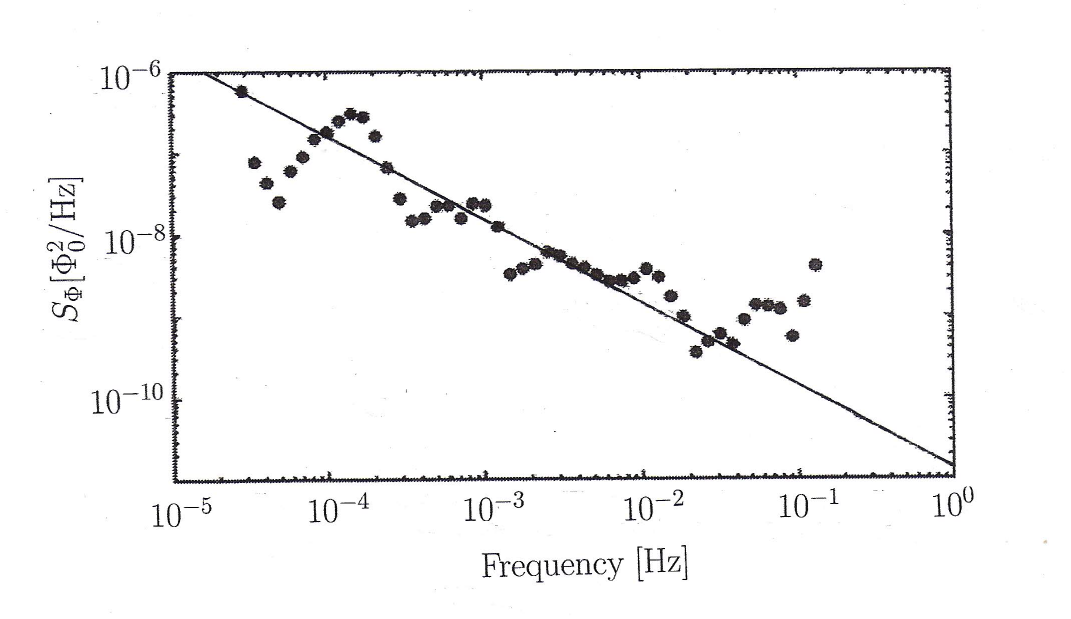}
\caption{{\bf Low-frequency flux noise power spectrum} as
measured by Bialczak et al. \cite{martinis2},
and comparison with the theoretical prediction eq.~(\ref{spectrum}) with $\theta_1^2=\bar{\theta_1^2}=\pi^2/3$
and $\alpha=2-n_s\approx 1.04$ (straight line).}
\end{figure}
Excellent agreement is found.
Note that no parameters are fitted, the theoretical prediction is just as it is, and it agrees perfectly
with the data.

The above measurements did not cover frequencies larger than $10^{-1}$ Hz.
In another experiment conducted by Sendelbach et al.\ \cite{sendel}, a
higher frequency region was probed, these data are displayed
in Fig.~2.
\begin{figure}
\includegraphics[width=6cm]{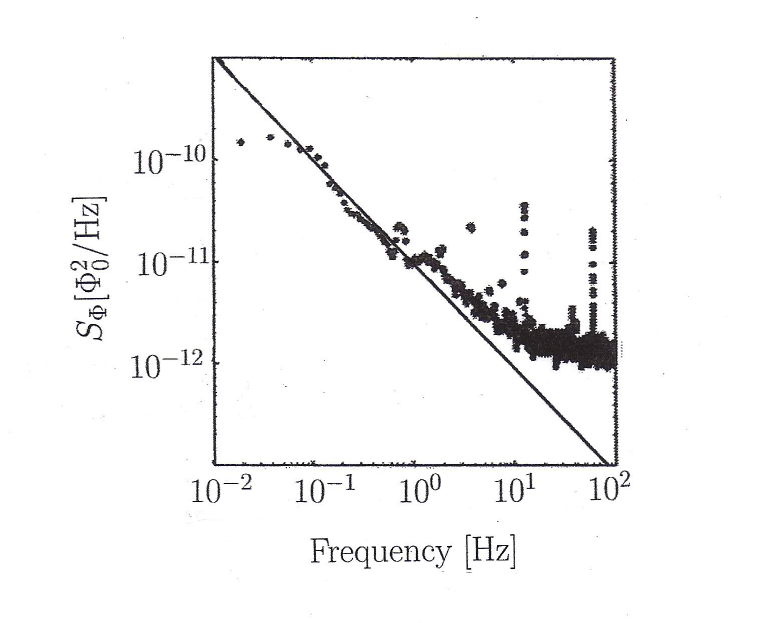}
\caption{{\bf Flux noise power spectrum
in the region 0.01...100 Hz} as measured
by Sendelbach et al. \cite{sendel}
and comparison with the theoretical prediction eq.~(\ref{spectrum}) with $\theta_1^2=\bar{\theta_1^2}=\pi^2/3$
and $\alpha =2-n_s\approx 1.04$ (straight line).}
\end{figure}
Again our theoretical prediction (\ref{spectrum}) agrees very well with the data in
the low-frequency region $10^{-1}...10^{0}$ Hz. For frequencies larger than about 1 Hz,
it is well-known (and verified in Fig.~2) that the noise spectrum becomes flatter, leading effectively
to $1/f^\alpha$ noise with $\alpha <1$, see e.g. \cite{clarke} for
recent very detailed measurements in this frequency region. In this region
secondary (non-universal) effects such as random flips of impurities in the surface material become important,
see \cite{koch} for suitable models in this direction.
Cosmological flux noise can still trigger
these complex internal surface processes, leading e.g.\ to the formation of clusters of surface spins.
However, in its original form the cosmological flux noise
is most dominant in the region $f<<1$ Hz,
where it can be identified by generating an exponent $\alpha =2-n_s \approx 1.04 >1$.

Sank et al.
\cite{martinis1} have recently performed a new series of high precision flux noise measurements with qubits
testing the frequency region $f=10^{-4}...10^{-1}$ Hz. These measurements are the most precise ones
currently available. With the new measurement technique described in \cite{martinis1} the fluctuations in the
measured noise spectra have become smaller. These
recent data are displayed in Fig.~3.
\begin{figure}
\includegraphics[width=7cm]{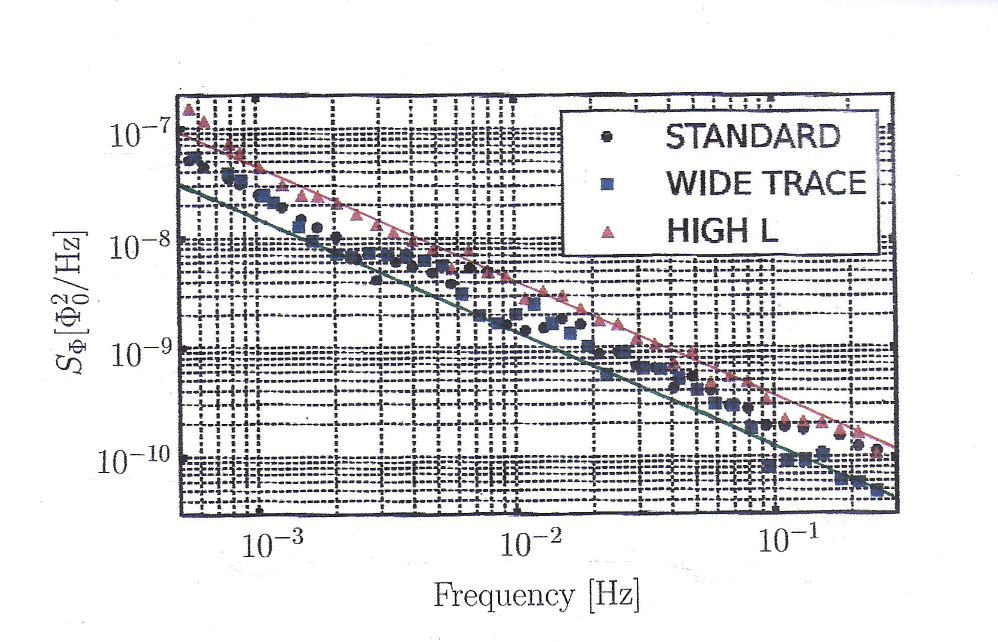}
\caption{{\bf Precision measurements of Sank et al. \cite{martinis1}} of the low-frequency flux noise power spectrum
and comparison with the theoretical prediction eq.~(\ref{spectrum}). The green line corresponds
to the cosmological average value
$\theta_1 =(\bar{\theta_1^2})^{1/2}= \pi /\sqrt{3}$, the red line to the maximum possible value $\theta_1=\theta_{max}=\pi$. All experimental
data lie between both lines and have the predicted slope 1.04.}
\end{figure}
Sank et al. report
a minimum flux noise strength of $3.5 \cdot 10^{-6}$ if extrapolated to
1Hz. This minimum value is in very good agreement with our theoretical prediction
of flux noise strength as given by eq.~(\ref{888}). The data of Sank et al.
can be
used to estimate the value of the initial misalignment
angle $\theta_1$ without any theoretical bias of what it should be. Using
eq.~(\ref{13}), we obtain from fitting the standard and wide trace data
the value $\theta_1=2.1\pm 0.4$.

It is interesting to compare the measured $1/f^\alpha$ flux noise
intensity from various recent experiments and to extract from this
the measured value of $\theta_1$.
As said before, the pure cosmological $1/f^\alpha$ flux noise is characterized by an exponent
$\alpha \approx 1.04$, whereas experimental data with an exponent significantly lower than 1 point towards
secondary effects, i.e.\ flux noise generated by surface
impurities and other material-dependent effects.
Hence, in Tab.~1 we restricted ourselves to experiments where the exponent
$\alpha$ was measured to be close to 1 (that is, flux noise data with, say, $\alpha \approx 0.6$ were ignored whereas data with $|\alpha -1.04|<0.2$ did enter
our analysis). The result of our analysis in Tab.~1
is the average value $\delta \Phi /\Phi_0=(3.84 \pm 0.96) \cdot 10^{-6}$ at 1Hz,
equivalent to $\theta_1 =2.04 \pm 0.67$. Within the error bars, this value
is compatible with the cosmological average value $\theta_1=\sqrt{\bar \theta_1^2}=1.81$.

\begin{table}[t]
{\tiny
\begin{center}
\begin{tabular}{| l ||  c |  c |}
\hline
Experiment       &$\frac{\delta \Phi}{\Phi_0}|_{f=1Hz}$    & Remarks \\
\hline \hline
Sank et al. \cite{martinis2}      & $3.9 \cdot 10^{-6}$ &  standard and wide trace \\
Sank et al. \cite{martinis2}    & $5.5 \cdot 10^{-6}$ &  high L, extrapolated to 1Hz  \\
Sendelbach et al. \cite{sendel} & $3.5 \cdot 10^{-6}$  &  direct measurement at 1Hz \\
Bialczak et al. \cite{martinis1}     & $4.0 \cdot 10^{-6}$  &  extrapolated to 1 Hz\\
Anton et al. \cite{anton}  & $3.5 \cdot 10^{-6}$ & direct measurement at 1Hz \\
Anton et al. \cite{clarke} & $4.4 \cdot 10^{-6}$ & direct measurement at 1 Hz \\
Bylander et al. \cite{bylander} & $2.1\cdot 10^{-6}$ & extrapolated from 1MHz to 1Hz \\
\hline \hline
average & $(3.84 \pm 0.96) \cdot 10^{-6}$ &  sample condition $|\alpha -1.04|<0.2$\\
\hline
\end{tabular}
\end{center}}
\caption{{\bf Flux noise strength at frequency 1Hz} as measured in different recent experiments \cite{martinis1,martinis2,sendel,anton,clarke,bylander}.}
\end{table}

\section*{Discussion}

Despite intensive research in the past
30 years \cite{koch-first}--\cite{wang} the deeper reason for the occurrence of $1/f^\alpha$ flux noise in qubits and SQUIDs
is still far from being fully understood, in particular in
the low-frequency region $f<1$ Hz. We have shown that
cosmologically generated flux noise
due to a cosmological field background surrounding the Earth can provide a suitable
explanation.
%
The predicted form of the spectrum is
in excellent agreement with the recent experimental observations of \cite{martinis1,martinis2,sendel};
this agreement is achieved without fitting any parameters.
In fact the only relevant parameter involved
for the cosmological flux noise is the
initial misalignment angle $\theta_1\in [-\pi, \pi]$.

As shown in this paper, $\theta_1$  can be extracted from precision measurements of the
flux noise intensity. The results of the various experimental groups
\cite{martinis1}--\cite{bylander} point to a value $\theta_1 \approx 2.04 \pm 0.67$,
compatible with the cosmologically expected average value 1.8.
We propose that
future systematic experimental tests should aim to separate universal from
non-universal (material and device dependent) effects. The universal low-frequency part of the flux noise spectrum may open up
a new experimental window to measure
power spectra of primordial fluctuations,
to provide high-precision measurements
of $\theta_1$ and $n_s$, and
to ultimately confirm
the existence of
cosmologically generated flux noise, by systematically excluding other (less fundamental) sources.
In fact, these types of experiments
could open up a new interdisciplinary field of research which we might call `nano-cosmology'.

 If the physical interpretation given in this paper is correct,
 then, rather than being just a nuisance in electronic devices,
 magnetic flux noise
 appears to contain valuable information
 about the state of the universe at an extremely early time,
 basically looking back to conserved
 frozen-in quantum fluctuations that were generated at the end of the inflationary period.

\section*{Methods}

We will now describe the methods that
 lead to the theoretical  prediction (\ref{spectrum}) in detail.
First, we will show that a spatial scale-invariant spectrum of density perturbations
can generate temporal $1/f$ noise for an observer that moves through this
fluctuating background with constant
velocity. Then we discuss how a primordial power spectrum can be conserved to
the current time in terms of misalignment perturbations of a suitable frozen-in cosmological field.
As a side product, we show that the potential energy of
this frozen-in field can generate constant vacuum energy density
that is comparable in magnitude to the currently observed dark energy density in the universe. The coupling
of the misalignment fluctuations into Josephson junctions via the flux quantization
condition is then discussed in the final subsection.

\subsection*{$1/f$ noise from a scale-invariant spectrum of spatial density fluctuations}

Let us
quite generally discuss an environment of energy density $\rho$ that exhibits spatial
density fluctuations $\delta \rho$ described by the (spatial) power spectrum $P(k)$:
\begin{equation}
\left( \frac{\delta \rho}{\rho} \right)^2= \int P(k) d \log k. \label{111}
\end{equation}
Here we use the definition of power spectrum as used by cosmologists and astrophysicists
(which is slightly different from that used by statistical physicists).
The astrophysical power spectrum $P(k)$ as defined in eq.~(\ref{111}) is the variance of
the relative density fluctuations $\delta \rho/\rho$ per logarithmic interval
$d \log k$, where $k$ denotes the scale. To evaluate the power spectrum at a
particular scale $k_0$, by convention the borders of the integral in eq.~(\ref{111}) are chosen as
$k_0$ and $ek_0$. An equivalent way of writing eq.~(\ref{111})
is thus
\begin{equation}
\left( \frac{\delta \rho}{\rho} \right)^2= \int_{k_0}^{ek_0} P(k) \frac{1}{k} d k. \label{222}
\end{equation}
The famous Harizon-Zeldovich spectrum is given by
\begin{equation}
P_{HZ}(k)=const,
\end{equation}
i.e. it is scale invariant. The primordial power spectrum of curvature
fluctuations as measured by the Planck satellite is
\cite{planck-old,planck}
\begin{equation}
P(k)= \Delta_R^2 (k)= A_s \cdot \left( \frac{k}{k^*} \right)^{n_s -1} \label{444}
\end{equation}
with $A_s\approx 2.2 \cdot 10^{-9}, n_s\approx 0.96$.

Let us now consider an observer that moves with constant velocity $v$ through this environment
and let
$\rho (t)=\bar{\rho} +\delta \rho (t)$ be the local density surrounding the observer
at time $t$. Let us consider the dimensionless stochastic process $Y(t)$ given by
$Y(t)=\delta \rho (t) /\bar{\rho}$, where $\bar{\rho}$ denotes
the average density. From eq.~(\ref{222}) it follows that the stochastic process $Y(t)$ has the
temporal power spectrum
\begin{equation}
S_Y(f)=\frac{1}{f} \cdot P(k)|_{k=\frac{f}{v}}. \label{333}
\end{equation}
Here we use the definition of (temporal) power spectrum as used by statistical
physicists (which has the dimension of time), and $f$ denotes the frequency.

From eq.~(\ref{333}) one sees that a medium with a spatial Harrison-Zeldovich spectrum $P(k)$
of density fluctuations
generates a temporal $1/f$ noise for an observer that moves through it with constant
velocity. More generally,
spatial fluctuations with spectral index $n_s$ generate
temporal noise with a power-law spectrum of type $1/f^\alpha$, where $\alpha=2-n_s$.
Note that often one needs very strong assumptions (such as
a uniform distribution of relaxation times \cite{koch}) to construct a plausible
temporal model for the origin of $1/f$ noise.
Here we see that spatial density fluctuations that are nearly-scale
invariant provide a very natural way to generate near- $1/f$ noise.
.

\subsection*{Conservation of the primordial power spectrum to the current time by a frozen-in cosmological field}

Primordial density fluctuations $\delta \rho / \rho$
are imprinted on any light field that is present
during cosmological inflation \cite{infla1, infla2, infla3,natural1, natural2}.
By a light field we actually mean a near-massless scalar field with a mass much smaller than
that of the inflaton \cite{riotto}.
Assume there is such a light field during inflation which is a relict
from an early quantum gravity epoch. We write this field as $a =f_a \theta$, where $f_a$ is a large energy scale,
assumed to be of the order of the Planck scale, and $\theta \in [-\pi, \pi]$ is a dimensionless angle variable.
We have chosen the symbol $a$ for this field since it may for example be an axion-like field \cite{peccei,wilczek,weinberg,stadnik,graham,visinelli,hertzberg,prl2013,pdu}.

In the simplest case we may just assume a quadratic potential $V(a)=\frac{1}{2} m^2 a^2$, where $m$ is
the mass of the scalar field under consideration, with $m << f_a$. For the QCD axion, a candidate
for cold dark matter in the universe, $f_a \sim 10^{11}$ GeV, but we are actually thinking
here of a different field that is a relict from a quantum gravity epoch, for which $f_a$ is larger, of the order of magnitude
of the Planck scale $10^{19}$ GeV.
These types of axion-like fields with large $f_a$ are predicted in a number of quantum gravity and inflationary models \cite{natural1,natural2}.
The scale $f_a \sim 10^{19}$ GeV also occurs as a fundamental lattice spacing in
a quantum description of discrete geometries where fundamental fields obeying Fermi-Dirac and Bose-Einstein
statistics arise in a natural way out of complex quantum network manifolds \cite{boguna, bianconi}.

If the above light field $a=f_a \theta$ (which is not the inflaton but an additional light field arising out of a unified theory of quantum gravity) is present during cosmological inflation, then
quantum fluctuations during inflation produce spatial fluctuations $\delta a$ of that field given by
\cite{hertzberg,visinelli}
\begin{equation}
\delta a = f_a \delta \theta = \frac{H_I}{2\pi}. \label{1111}
\end{equation}
Here $H_I$ is the Hubble parameter during inflation. These field fluctuations correspond to density
fluctuations given by
\begin{equation}
\frac{\delta \rho_a}{\rho_a}=\frac{2}{\theta} \delta \theta =\frac{H_I}{\pi \theta f_a}\label{22222}
\end{equation}
since $\rho_a = \frac{1}{2} m^2 f_a^2 \theta^2$ and $\delta \rho_a =m^2 f_a^2 \theta \delta \theta$.
Spatial fluctuations in the energy density of this cosmological field can thus be equivalently regarded as
representing (encoding) fluctuations $\delta \theta$ of a misalignment angle $\theta$, as given by eq.~(\ref{22222}).
These angle perturbations are present on a huge range of scales, due to the
exponential expansion of the scale factor during inflation.

Let us now check the conditions under which these angle perturbations produced during inflation
can be conserved to the current time.
The equation of motion of the field $a$ in an expanding flat Friedmann-Robertson-Walker background is \cite{hertzberg}
\begin{equation}
\ddot{a} +3H \dot{a}- \frac{1}{R^2} \nabla^2 a +m^2 a =0, \label{field-eq}
\end{equation}
where $H$ is the (temperature dependent) Hubble parameter and $R$ the scale factor.
The spatial gradient terms in the above equation are very small and can be neglected.
We want this field to be frozen in up to the current time, in order
to conserve the primordial power spectrum, meaning the angle perturbations have not evolved at all so far.
This means the kinetic
energy $E_{kin}=\frac{1}{2} \dot{a}^2$ must still be much smaller than the
potential energy $E_{pot}=\frac{1}{2}m^2 a^2$. This condition of a frozen-in state is realized if the Hubble damping is still strong enough as compared to the potential strength, i.e. if
\begin{equation}
m \lesssim H_0 \label{1111}
\end{equation}
where $H_0$ is the Hubble parameter at the current time.

On the other hand, as mentioned before, we may assume that this cosmological field
is a relict from a quantum gravity epoch, i.e. an epoch
where possibly all interactions were in a unified state, and then this symmetry was broken. This requires that the energy scale $f_a$
(which corresponds to a symmetry breaking scale \cite{peccei}) should be
of the order of the Planck scale, or even higher:
\begin{equation}
f_a \gtrsim m_{Pl} \label{2222}
\end{equation}
Equations (\ref{1111}) and (\ref{2222}) imply that the mass parameter $m$ must be extremely small, $m \lesssim H_0$, and the energy scale $f_a \gtrsim m_{Pl}$
extremely large. Still the product $mf_a$ which enters into
the potential energy gives a well-defined finite value which has a physical interpretation, namely
we get potential energy that has the same order of magnitude as the currently
observed dark energy density $\rho_{dark}$ in the universe \cite{peebles,prd2004,planck-dark}:
\begin{equation}
E_{pot} = \rho_a \sim m^2 f_a^2 \sim H_0^2 m_{Pl}^2 \sim \rho_{dark}
\end{equation}
Indeed for a flat universe one has
\begin{equation}
H^2 = \frac{8}{3}\pi G (\rho_{dark}+\rho_r +\rho_m)
\end{equation}
and at the current time $(H=H_0)$ the dark energy density $\rho_{dark}$ is observed to dominate as compared to the radiation density
$\rho_r$ and matter density $\rho_m$. Note that in units where $\hbar=c=1$ we have $G=m_{Pl}^{-2}$.

Hence, as a by-product of our efforts to construct a light field that conserves the primordial power spectrum to the current time,
we have obtained dark energy. Dark energy could be identified with the constant potential energy
of the frozen-in cosmological field $a$. Since this field is static up to the current time,
the energy density does not evolve in time and represents a small cosmological constant.

Note that in contrast to the QCD axion dark matter field, which is initially frozen-in but leaves its
frozen-in (time-independent) state shortly before the QCD phase transition to start oscillating behaviour, we are here postulating a different
axion-like field which is still in a frozen-in state up to the current time. Its potential energy is not
given by QCD vacuum energy (as for the QCD axion) but by the dark energy density $\rho_{dark}$ in the
universe. For the simplest model, a cosmological constant $\Lambda$ and non-evolving dark energy density, this
energy density is given by $\rho_{dark}=\frac{c^2}{8\pi G} \Lambda$. The most recent Planck measurements \cite{planck},
based on the $\Lambda CDM$ model, yield the numerical value $\rho_{dark}=(3.35 \pm 0.16)$ GeV/$m^3$.

In the model proposed in this paper we associate the cosmological field $a$ with frozen-in magnetic flux fluctuations
associated with a cosmological constant, which are completely decoupled from the rest of the universe, and which do not
evolve in time (more complicated models with an evolving $\rho_{dark}$ can also be studied but are not subject of this paper).
It is interesting to check what typical values of magnetic field strength of this `dark' magnetic field $B_0$ one formally obtains if one assumes that
a fraction $\eta$ of the dark energy density $\rho_{dark}$ in the universe is actually frozen-in magnetic field energy $\rho_B$.
Writing $\rho_B= \frac{1}{2\mu_0}B_0^2= \eta \rho_{dark}$
one obtains
\begin{equation}
|B_0|=\sqrt{\eta} \sqrt{2\mu_0 \rho_{dark}}
\end{equation}
which for $\eta=1$ numerically evaluates to $|B_0|=(3.67\pm 0.08) \cdot 10^{-8}$ T
provided one uses for $\mu_0$ the usual magnetic permeability of the vacuum. This is a very small magnetic field, comparable in size
to small magnetic fields measured in the outer heliosphere. It is unlikely that the above
formal magnetic field $B_0$ can ever be measured, since it can point into any direction of space equally likely.
Still it is interesting to check what typical area $A^*$ one obtains if one writes down a flux quantization
condition of the form
\begin{equation}
B_0 A^*=n \Phi_0 =n \frac{h}{2e},
\end{equation}
where $n$ is an integer.
For the choice $n=1=\eta$ we obtain the numerical value $A^*=(5.64 \pm 0.12) \cdot 10^{-8}m^2$, which corresponds to
a length scale $r^*=\sqrt{A^*/\pi}=(134 \pm 2) \; \mu m$. This is of the same order of magnitude as
the loop radius of a typical (big) SQUID. It is encouraging that one does not get any exotic length scales
but parameters that make sense in SQUID physics.
While it is unlikely that the above formal magnetic field $B_0$ associated with dark energy
can ever be measured directly, our main proposal in this paper is that tiny fluctuations and inhomogenities
of the associated flux can be measured in a highly sensitive SQUID environment,
and lead to the experimentally observed flux noise. This will be worked out in more detail the following section.

\subsection*{Coupling of misalignment angle fluctuations into SQUIDs and qubits}
\label{C}

While the potential energy of the field $a$ is practically constant, and the field is very homogeneous, there are still tiny spatial density
fluctuations imprinted onto this nearly massless field, originating from quantum fluctuations during inflation.
These fluctuations
of the field $a$ are equivalent to tiny conserved spatial misalignment angle fluctuations, and they
should still have the same power spectrum as in the very early universe.

Let us now discuss a possible mechanism how a fluctuation of the misalignent angle
surrounding locally the moving Earth
can couple into Josephson junctions, SQUIDs or qubits.
Let us first consider standard SQUID physics.
If two Josephson junctions, one described by the gauge-invariant phase difference $\varphi_1$ and the other one
by the gauge-invariant phase difference $\varphi_2$ form a SQUID (superconducting quantum interference
device), then it is well-known that the difference
$\varphi_1 - \varphi_2$ satisfies \cite{tinkham}
\begin{equation}
\varphi_1 -\varphi_2 = 2\pi \frac{\Phi}{\Phi_0} \mod 2\pi. \label{1}
\end{equation}
Here $\Phi$ is the magnetic flux included in a closed loop containing the weak link region of the SQUID,
and $\Phi_0=h/2e$ denotes the flux quantum. Eq.~(\ref{1}) is a simple consequence
of the fact that the joint macroscopic wave function describing the physics of both junctions
forming the SQUID must be unique
\cite{tinkham}.

From the above it is obvious that an uncertainty $\delta \varphi$ in the phase difference
$\varphi_1 - \varphi_2$ can be equivalently regarded as a flux uncertainty $\delta \Phi$:
\begin{equation}
\delta \varphi = 2\pi \frac{\delta \Phi}{\Phi_0}. \label{paul}
\end{equation}
Fluctuations in (or uncertainties in the knowledge of) the angle variable of a Josephson junction thus imply magnetic flux noise.

In analogy to this, in \cite{prl2013} it was proposed that phase differences in Josephson junctions are influenced by
phase differences of a surrounding axion condensate, in the sense that any change $\delta \theta$ in the
surrounding axion condensate is accompanied (or compensated) by a mirror change $\delta \varphi$
of the electromagnetic phase difference in the corresponding Josephson junction,
\begin{equation}
\delta \varphi = \delta \theta. \label{equality}
\end{equation}
The physical meaning of eq.~(\ref{equality}) is that the axion field
sets the background to which all Josephson phases need to be related. If the background
changes, so does the Josephson phase.
As the Earth
moves through a spatially inhomogeneous axion background, the axion misalignment angle exhibits tiny changes
$\delta \theta$ which are accompanied by a corresponding mirror change of the electromagnetic phase variable in the junction.
From eq.~(\ref{equality})
we get
\begin{equation}
\delta \varphi (t) = \delta \theta (\vec{x} (t)), \label{12}
\end{equation}
where $\vec{x}(t)$ is the position of the Josephson junction on the Earth moving relative
to the background field of spatial misaligment angle fluctuations.

While for a more detailed discussion of the underlying mathematics we refer to \cite{prl2013,pdu}, let us here give a simple physical
argument why a SQUID-like interaction of the form (\ref{equality}) is the only consistent way to introduce a coupling between
axion fields and Josephson junctions. Axions are described by a cosine potential
$V(a)=m^2 f_a^2 (1-\cos \theta)$ in the angle variable $\theta =a/f_a$, and the physical effect of any perturbation $\delta \theta$
of the angle must be invariant under the
transformation $\delta \theta \to \delta \theta +2 \pi$.
Moreover, also SQUID physics is invariant under the
transformation $\delta \varphi \to \delta \varphi +2 \pi$, as only
the phase modulo $2\pi$ of the macroscopic wave function matters. Whatever the interaction
between SQUIDs and axions, performing both
transformations simultaneously should not change the physics. If we
assume a linearized relation of the form
\begin{equation}
\delta \varphi = C \delta \theta
\end{equation}
with some unknown coupling constant $C$,
then any physics should be invariant under the above transformations of increasing the angle perturbations
by $2 \pi$ on either side. Hence
\begin{equation}
\delta \varphi +2\pi = C(\delta \theta +2\pi)= C\delta \theta +2\pi C
\end{equation}
Since $\delta \varphi=C \delta \theta$ we thus obtain
\begin{equation}
C=1
\end{equation}
which proves eq.~(\ref{equality}).

Combining eqs.~(\ref{12}), (\ref{paul})
and (\ref{22222}), we get a concrete prediction for the flux noise generated by the background misalignment fluctuations:
\begin{equation}
\frac{\delta \Phi}{\Phi_0} =\frac{1}{2\pi} \delta \varphi =\frac{1}{2\pi} \delta \theta = \frac{\theta_1}{4 \pi} \frac{\delta \rho_a}{\rho_a}. \label{777}
\end{equation}
Here $\theta_1\in [-\pi , \pi]$ denotes the initial value of the cosmological field angle.
Using also eq.~(\ref{333}) we end up with
\begin{equation}
S_\Phi (f) = \frac{\theta_1^2}{16 \pi^2} \Phi_0^2 P(k)|_{k=f/v} \frac{1}{f}. \label{spectrum-gen}
\end{equation}
In particular, the primordial power spectrum of scalar perturbations (\ref{444})
yields the prediction
\begin{equation}
S_\Phi (f) = \frac{\theta_1^2\Phi_0^2 A_s}{16 \pi^2}\left( \frac{f}{vk^*} \right)^{n_s-1} \frac{1}{f}. \label{spectrum-again}
\end{equation}
which is the main result of this paper.

Our derivation of eq.~(\ref{spectrum-again}) was based on the assumption of a simple quadratic potential $V(a)$ for
the cosmological field $a$. More general, for a given arbitrary potential $V(a)$ one obtains the more general result
that the effective coupling parameter $\theta_1$ in eq.~(\ref{spectrum-gen}) and (\ref{spectrum-again}) is given by
\begin{equation}
\theta_1= \frac{2 V(a)}{f_a V'(a)}
\end{equation}
Here $a=a(0)=f_a\theta (0)$ denotes the initial field value of the cosmological field $a$.
For example, for a cosine potential $V(a)=m^2f_a^2(1-\cos \theta)$ one obtains
\begin{equation}
\theta_1 = 2 \frac{1-\cos \theta (0)}{\sin \theta (0)}.
\end{equation}
$\theta (0)=\frac{\pi}{\sqrt{3}} \approx 1.81$ implies $\theta_1=2.55$, still compatible with the flux noise measurements
in Table~1.


\subsection*{Acknowledgements}

C.B.'s research is supported by EPSRC under grant No. EP/N013492/1.

\subsection*{Author Contributions}

C.B. did the research and wrote the manuscript.

\subsection*{Additional Information}

{\bf Competing financial interests:} The author declares no competing financial interests.

\end{document}